\documentstyle [preprint,aps,psfig] {revtex}

\begin{document}

\title{Longitudinal solitons in carbon nanotubes }
\author{T.Yu.Astakhova$^a$, O.D.Gurin$^a$, M.Menon$^{b,c}$\cite{email1},
and G.A.Vinogradov$^a$\cite{email2}}

\address{$^a$ Institute of Biochemical Physics RAS, ul.Kosygina 4,
Moscow
117334, Russia}
\address{$^b$Department of Physics and Astronomy, University of
Kentucky, Lexington, KY 40506-0055, USA}
\address{$^c$Center for Computational Sciences, University of
Kentucky, Lexington, KY 40506-0045, USA}

\maketitle

\begin{abstract}
We present the results on the soliton excitations in carbon nanotubes
(CNT) using Brenner's many-body potential. Our numerical simulations
demonstrate high soliton stability in $(10,10)$ CNT. The interactions
of solitons and solitary excitations with CNT defects are found to be
inelastic if the excitations and defects length scales are comparable,
resulting in a substantial part of soliton energy being distributed
inhomogeneously over the defect bonds. In these solitary--cap
collisions the local energy of few bonds in the cap can exceeds the
average energy by an order of magnitude and more. This phenomenon
denoted as ``Tsunami effect'' can contribute dynamically to the
recently proposed ``kinky chemistry''. We also present results of
changes in the local density of states and variations in the atomic
partial charges estimated at different time instants of the solitary
Tsunami at the nanotube cap.
\end{abstract}

\section{Introduction}

Since the discovery of carbon nanotubes by Iijima,\cite{ijima} a rich
variety of carbon nanotube morphologies have been experimentally
observed. Carbon nanotubes (CNT) are very intriguing objects for both
experimental and theoretical investigations due to the many unusual
properties they exhibit. Their potential for practical applications,
especially in the area of nanotechnology, is very promising. Among
their many uses envisioned include applications in nanoscale electronic
devices, field-emission displays and quantum wires.\cite{nano}
Nanotube doping and structure modification result in the formation of
heterojunctions,\cite{heteroj} diodes,\cite{diode} quantum
dots,\cite{quantumdot} field effect transistors \cite{fieldtrans} and
one--electron conductors.\cite{oneelectr}

The dynamical properties of carbon nanotubes are of great interest due
to their potential for useful practical applications.\cite{mechanics}
The theoretical methods for explaining these properties range from
accurate {\sl ab initio} methods to approximate empirical schemes.
Few of experimental findings, however, were analyzed using empirical
methods that take into account harmonic contributions such as effects
of different vibrational modes, explicitly. In particular, nonlinear
localized excitations can transfer energy and be involved in various
processes of interest. There have been some speculations on the role
played by solitary excitations in heat transfer, polymer destruction
and other processes occurring in molecular systems.\cite{soliton}
While in most cases of interest harmonic approximation is sufficient
to describe molecular processes, some molecular materials contain
atoms interacting through nonlinear inter-atomic potentials which can
give rise to soliton excitations. The investigation of these
nonlinear effects is, therefore, interesting and timely.

In this work we investigate the nonlinear effects in carbon nanotubes
giving rise to solitons using empirical methods. We employ Brenner's
nonlocal many-body potential for carbon systems.\cite{Brenner} Its
simple analytical form as well as its reliability for carbon systems
coupled with its computational efficiency makes it a natural candidate
for such investigations. The nanotube chosen for our simulations is a
$(10,10)$ tube containing 1\,000-11\,000 atoms. The simulations were
performed for free, rigid and capped tube ends. Our investigations
include soliton stability, reflection from free and capped CNT ends,
solitons collisions as well as the influence of thermal fluctuations.
We find that the interaction of solitary excitations with the CNT cap
can be either elastic or inelastic. In the latter case some part of
the energy is retained as an internal excitation of the cap with the
remaining portion distributed highly inhomogeneously over the C-C
bonds. This effect is found to be most pronounced for the non
symmetrical caps. The degree of energy concentration depends
sensitively on the cap structure and solitary energy and profile. In
some cases the energy of few bonds in the cap exceeds the average
solitary energy by an order of magnitude or more. Similar phenomenon
is also observed for other defects in CNTs.

The paper is organized as follows: In Sec. \ref{nonlinear} we derive a
nonlinear expansion of the Brenner's potential for longitudinal and
radial atomic displacements. Sec. \ref{korteweg} contains the
formulation of an analytical equation for the longitudinal
displacements of atoms along the CNT axis. In this section we derive
the one--dimensional Korteweg -- de~Vries equation assuming angular
homogeneity and neglecting the radial displacements and obtain the
corresponding supersonic soliton solution. Results of the numerical
simulations of soliton behavior in CNTs are given in Sec.
\ref{numerical}. Sec. \ref{inter} contains detailed analysis of
solitary collisions with CNT defects. The possible soliton
contributions to various chemical and physical properties of CNT are
briefly discussed in Sec. \ref{concl} with special attention paid to
the changes in the local density of states and variations in partial
atomic charges at different instances of the soliton--cap interaction.
Along with summarizing our results, we also propose possible ways of
soliton generation in this section.

\section{Nonlinear expansion of the Brenner's potential}
\label{nonlinear}

Carbon nanotubes (only single--walled CNT are considered in this work)
are highly anisotropic and ordered objects with great aspect ratio.
The CNT surface is formed by a graphitic sheet folded into a cylinder
with bond lengths and angles differing slightly from graphite on
account of the strain induced by the folding. The inter-atomic C-C
potential is highly unharmonic and can be approximated using classical
many-body potentials containing either exponential or power terms.
Generally, these potentials do not provide an exactly solvable
dynamical equation in the nonlinear continuum approximation and,
therefore, some approximations needs to be made in order to yield
desired solutions.

The unharmonic solutions in nonlinear 1D lattices are frequently
approximated by the Korteweg -- de~Vries (KdV) equation.\cite{x^3}
Since CNTs are quasi--1D objects, in order to obtain nonlinear
excitations in CNT, it is reasonable to approximate CNT as a 1D
lattice with nonlinear effective potential. We then make the
continuum approximation for this system and derive the
corresponding KdV equation.

We obtain the nonlinear effective potential for CNT by expanding
Brenner's many body-potential in Taylor series up to the third order
terms. The corresponding Newtonian equations of motion in finite
differences are then derived and finally, the continuous analogue
(nonlinear partial differential equation) for the system is obtained.
This results in the KdV equation which describes longitudinal
nonlinear excitations (solitons) in CNT.

According to Brenner's many-body potential,\cite{Brenner} the bond
energy
between adjacent atoms $i$ and $j$ is expressed as,
\begin{equation}
\label{Brenner}
E_{ij}^b = V_{ij}^R - \overline{B}_{ij}\,V_{ij}^A.
\label{eq1}
\end{equation}
where $V_{ij}^R$ and $V_{ij}^A$ are, respectively, exponential
repulsive and attractive terms: $V^R(r_{ij}) =
27.27\,exp[-3.28\,(r_{ij} -1.39)], V^A(r_{ij}) =
33.27\,exp[-2.69\,(r_{ij} -1.39)]$. $\overline{B}_{ij}$ represents an
environment dependent many--body coupling between atoms $i$ and $j$
containing geometric information associated with the
system.\cite{Brenner} The total energy of the system is obtained by
summing Eqn. \ref{eq1} over all bonds. The energy and distances are
measured in eV and {\AA}ngstr{\"o}ms, respectively.

The reliability of Brenner's potential has been demonstrated in
many numerical calculations of carbon systems in different
phases (graphite, diamond, fullerenes and large clusters),
in both ground state as well as in non-equilibrium states including
chemical reactions.\cite{B_application}
Furthermore, the simple form of the potential allows
us to easily derive analytical expressions for forces. This also
allows efficient large scale simulations of the system using
available computer resources.

\begin{figure}
\hspace{-1cm}
\centerline{\psfig{file=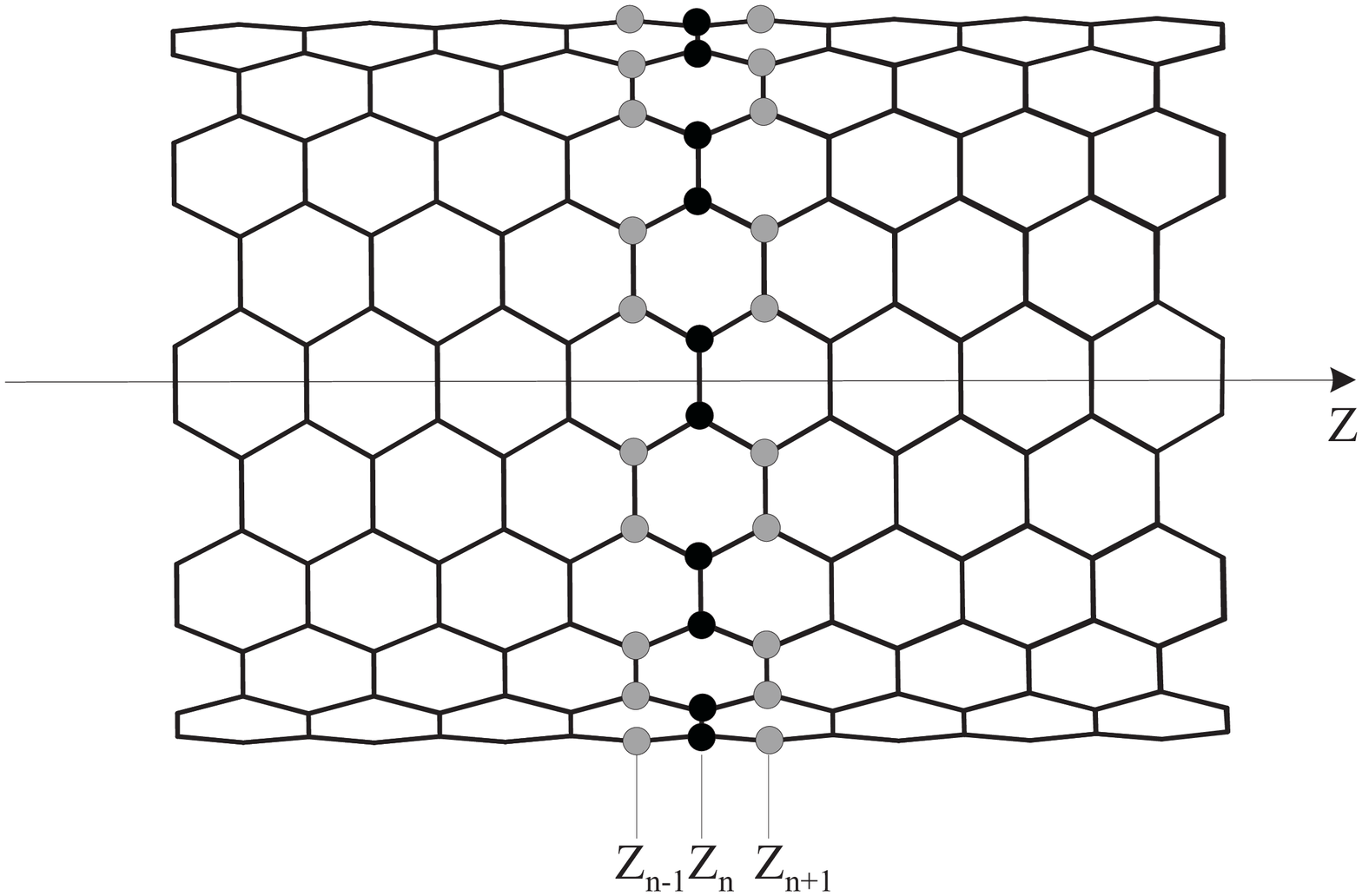,angle=0,width=8cm}}
\caption{A part of $(10,10)$ nanotube. The tube is oriented along
the $z$-axis. Atoms in $n, (n \pm 1)$--th layers are marked by black
and gray circles, respectively.}
\label{soliton_fig1}
\end{figure}

In the present work we consider only armchair $(m,m)$ CNTs. The
procedure and results are similar in the case of zigzag $(m,0)$
nanotubes. We use the cylindrical coordinate system ($R, \Phi, Z$) for
obvious reasons and align the tube axis along z--axis (Fig.
\ref{soliton_fig1}). With this choice there are $2\,m$ atoms in the
CNT layer having the same $z$--coordinates. The equilibrium distance
between layers, $\ell _0$, has a slight dependence on the CNT diameter
(see Table 1). For a (10,10) tube, $\ell _0 \approx 1.26 \AA$.

Let us consider a cylindrically symmetrical disturbance in the CNT
geometry in which all 2$m$ atoms in the $n$--th layer have identical
$z$-- and radial displacements from their equilibrium positions
$Z^0_n$ and $R^0$: $Z_n = Z^0_n + \zeta_n$ and $R_n = R^0 + \rho_n$,
where $Z_n$ and $R_n$ are the perturbed coordinates and $\zeta_n$ and
$\rho_n$ are the displacements from the equilibrium. Then, the
coordinates of $i$--th atom in $n$--th layer are $(R^0 +\rho_n, \,
\Phi ^0_i, \, Z^0_n+\zeta_n)$. If $\zeta$ and $\rho $ are much smaller
when compared to characteristic length scales ($\zeta_n \ll \ell _0$
and $\rho_n \ll R^0$), then the interatomic potential
(Eqn. \ref{Brenner}) can be expanded in the Taylor series and the
terms up to the third order retained:
\begin{equation}
\label{Bren_tay1}
E = E_0 + \sum_{n = 1}^{N_{\ell}} E_n,
\label{eq2}
\end{equation}
where $E_0$ is the ground state energy of the relaxed CNT and the
perturbation energy is the sum over all $N_{\ell}$ layers of CNT. For
every layer $E_n = 2\,m\,E_n^a$ is the sum of perturbation energies of
$2\,m$ atoms in the $n$--th layer due to atomic displacements, and
energy of an atom $E_n^a =1/2\,(\, E_1 + E_2 +E_3 \,)$, where $E_1$,
$E_2$ and $E_3$ are energies of bonds emerging from any atom in the
$n$--th layer (see Fig. \ref{soliton_fig1}). It is convenient to
reduce both $\zeta $ and $\rho $ to dimensionless units: $\zeta_n \to
\zeta_n/\ell _0, \,\, \rho_n \to \rho_n/\ell _0$.

Then an expansion $E_n^a$ over atomic $\zeta $ and $\rho $
displacements in the $n$--th layer has a form:
\begin{equation}
\label{Bren_tay3}
\begin{array}{ll}
E_n^a =

& \displaystyle\frac{ a_1}{2}\,(\zeta_{n+1} - \zeta_n)^2
- \displaystyle\frac{ a_2}{3}\,(\zeta_{n+1} - \zeta_n)^3 \\

& + \displaystyle\frac{b_1}{2}\,\rho_n^2
+ \displaystyle\frac{b_2}{2}\,(\rho_{n+1} + \rho_n)^2
+ \displaystyle\frac{b_3}{3}\,(\rho_{n+1} - \rho_n)^2\,(\rho_{n+1} +
\rho_n)
- \displaystyle\frac{b_4}{3}\,\rho_n^3 \\

& + \displaystyle\frac{c_1}{2}\,(\zeta_{n+1} - \zeta_n)\,(\rho_{n+1} +
\rho_n)
- \displaystyle\frac{c_2}{3}\,(\zeta_{n+1} - \zeta_n)^2\,(\rho_{n+1} +
\rho_n)
+ \displaystyle\frac{c_3}{3}\,(\zeta_{n+1} - \zeta_n)\,(\rho_{n+1} -
\rho_n)^2.

\end{array}
\end{equation}
Numerical values for the expansion coefficients in Eqn.
\ref{Bren_tay3} are given in Table I for nanotubes $(5,5), (10,10),
(15,15)$ and $(20,20)$. All coefficients in Eqn. \ref{Bren_tay3} are
measured in eV and variables $\zeta$ and $\rho$ are chosen to be
dimensionless. Here only the leading terms in the nearest--neighbor
interaction are retained. In this assumption the CNT is treated as a
one--dimensional lattice with $2\,m$ carbon atoms at each lattice site
interacting through an effective potential given by
Eqn. \ref{Bren_tay3}.

Terms in Eqn. \ref{Bren_tay3} can be categorized in three groups. The
first contains only longitudinal variables $\zeta$, the second
contains only the radial variables $\rho$, and the third contains
cross terms. The coefficients $a_1, a_2$ in the first group have the
largest absolute values and are mostly independent on the CNT diameter
(see Table I). The coefficients in other groups ($b_i$ and $c_i $)
have smaller numerical values but strong dependence on the CNT
diameter: the larger the CNT diameter, the smaller the values of these
coefficients. This indicates that the contribution from the
longitudinal degree of freedom to the dynamical properties of CNT is
dominant, and the role of other members decreases with the increase in
the CNT diameters. We, therefore, consider only the longitudinal
degree of freedom in the study of nonlinear dynamics. Analytical and
numerical results for the case in which both longitudinal and radial
degrees of freedom are considered will follow in a future report. We
note that these results differ only in minor qualitative details, the
main features being determined by the longitudinal degree of freedom.

Note that the oscillatory equation for the radial vibrations
$$
M \, \displaystyle\frac{\partial^2 \rho_n}{\partial t^2} =
- b_1\, \rho _n
$$
gives the lowest--frequency ``breathing'' mode in harmonic radial
approximation for CNTs, and the Raman frequency for this mode for the
$(10,10)$ CNT is $\approx 165$ cm$^{-1}$, where $M$ is the mass of
carbon atom.

The Newtonian equation of motion for longitudinal $\zeta
$--displacements (neglecting radial degrees of freedom) is:
\begin{equation}
\label{Newton_z}
\begin{array}{ll}
M \, \displaystyle\frac{\partial^2 \zeta _n}{\partial t^2} =
& a_1 \,[\,(\zeta_{n+1} - \zeta_n) - (\zeta_{n} - \zeta_{n-1})\,]
- a_2 \,[\,(\zeta_{n+1} - \zeta_n)^2 - (\zeta_{n} - \zeta_{n-1})^2\,].
\end{array}
\end{equation}
This is a nonlinear equation in finite differences. In the next
section we derive its continuum analog and examine the consequences.

\section{Korteweg -- de Vries solitons in CNT}
\label{korteweg}

Eqn. \ref{Newton_z} presents a true 1D nonlinear problem with only one
variable $\zeta $. The corresponding continuum equation is obtained
using the work of M.Toda.\cite{Toda} Let the relative displacement of
all atoms in $n$--th and $(n-1)$--th layers be given by $\chi _n =
\zeta_{n} - \zeta_{n-1}$. The continuum approximation is then valid if
only the excitations with wave lengths much greater than the lattice
constant $\ell _0$ are allowed. The expression for the 1D expansion
of Brenner's potential (Eqn. \ref{Bren_tay1}) in Taylor series can be
written as:
\begin{equation}
\label{discrete}
E = E_0 + \sum_{n = 1}^{N_{\ell}}
\left(\frac{a_1}{2}\,\chi _n^2 - \frac{a_2}{3}\,\chi _n^3\right).
\end{equation}
The corresponding equation of motions for relative displacements
$\chi$, is:
\begin{equation}
\label{chi_1}
M \, \displaystyle\frac{\partial^2 \chi_n}{\partial t^2} =
2\,\left[a_2\,\chi_n^2 - a_1\,\chi_n \right]
- \left[a_2\,\chi_{n+1}^2 - a_1\,\chi_{n+1} \right]
- \left[a_2\,\chi_{n-1}^2 - a_1\,\chi_{n-1} \right].
\end{equation}

Eqn. \ref{chi_1} contains terms with $\chi _{n \pm 1}$ and $\chi _{n
\pm 1}^2$. It is convenient to introduce a shift operator ${\cal
P}^{\pm 1} = \exp \left (\pm \displaystyle\frac{\partial}{\partial
n}\right )$ for a shift along the discrete chain by one step to the
right (left) such that ${\cal P}^{\pm 1}\,f(n) = f(n \pm 1)$, where
$f(n)$ is an arbitrary function of discrete coordinate $n$.
Eqn. \ref{chi_1} can then be written as:
\begin{equation}
\label{chi_2}
M \, \displaystyle\frac{\partial^2 \chi_n}{\partial t^2} =
\left[ 2 - \exp{(\frac{\partial}{\partial n})} -
\exp{(-\frac{\partial}{\partial n})} \right]
\left( a_2\,\chi_n^2 - a_1\,\chi_n \right),
\end{equation}
or
\begin{equation}
\label{chi_3}
M \, \displaystyle\frac{\partial^2 \chi_n}{\partial t^2} =
- \left[2 \,
\sinh \displaystyle{ \left(\frac{1}{2} \frac{\partial}
{\partial n}\right )}
\right]^2
\left( a_2\,\chi_n^2 - a_1\,\chi_n \right),
\end{equation}
\noindent
resulting in,
\begin{equation}
\label{motion}
M \, \frac{\partial ^2 \chi }{\partial t^2} -
a_1\,\left[ 2\sinh \left( \frac{1}{2} \frac{\partial}{\partial n}
\right)\right]^2 \chi + a_2\, \left[ 2\sinh \left(\frac{1}{2}
\frac{\partial}{\partial n} \right)\right] ^2 \chi ^ 2 = 0.
\end{equation}

If $\sinh \displaystyle{ \left(\frac{1}{2} \frac{\partial} {\partial
n}\right )}$ in the third term in Eqn. \ref{motion} is expanded in
Taylor series up to the linear term, then the equation can be
rewritten as
\begin{equation}
\label{motion1}
M \, \frac{\partial ^2 \chi }{\partial t^2} - a_1\,\left[ 2 \sinh
\left( \frac{1}{2} \frac{\partial}{\partial n}\right) \right] ^2 \chi
+ a_2\,\frac{\partial ^2}{\partial n^2}\, \chi ^ 2 = 0,
\end{equation}
and can be factored into the following products:
\begin{equation}
\label{motion2}
\left\{ \sqrt{M} \frac{\partial}{\partial t} \mp
2\sqrt{a_1}\sinh\left (\frac{1}{2} \frac{\partial}{\partial n}\right )
\pm \frac{a_2}{\sqrt{a_1}} \frac{\partial}{\partial n} \right\} \left\{
\sqrt{M} \frac{\partial}{\partial t} \pm 2\sqrt{a_1}\sinh\left
(\frac{1}{2} \frac{\partial}{\partial n}\right ) \mp
\frac{a_2}{\sqrt{a_1}} \chi \frac{\partial}{\partial n} \right\}
\chi = 0.
\end{equation}

Eqns.\,(\ref{motion1}) and (\ref{motion2}) are equivalent modulo
terms of third order. Eqn.\,(\ref{motion2}) has a form of a product of
two operators $\widehat F_1 \, \widehat F_2 \, \chi$.
If we make a further assumption: $ \widehat F_2 \, \chi = 0$,
\cite{Toda}
then the wave packet moving to the right (left) is described by
\begin{equation}
\label{motion_right}
\left\{ \sqrt{M} \, \frac{\partial}{\partial t} \pm
2\,\sqrt{a_1}\sinh \left(\frac{1}{2} \frac{\partial}{\partial n}\right)
\mp \frac{a_2}{\sqrt{a_1}} \, \chi \, \frac{\partial}{\partial n}
\right\} \chi = 0.
\end{equation}

If $\sinh\displaystyle{\left (\frac{1}{2}
\frac{\partial}{\partial n}\right )}$ in (\ref{motion_right})
is further expanded in Taylor series up to the third order terms and $n$
is substituted by continuum variable $z$, one gets the corresponding
continuum equation:
\noindent
\begin{equation}
\label{contin}
\frac{\partial{\chi}}{\partial{t}} +
k_1 \, \frac{\partial{\chi}}{\partial{z}} +
k_2 \, \frac{\partial ^3\chi}{\partial{z^3}} -
k_3 \, \chi \,\frac{\partial{\chi}}{\partial{z}} = 0,
\end{equation}
where
$k_1 = k, \, k_2 = k/24, \,
k_3 = \displaystyle{\frac{a_2}{a_1}}\,k, \,$
and $k= \pm \sqrt{a_1/M}$.

Eqn.(\ref{contin}) can be reduced to the standard KdV equation
$(u_{\tau} + 6uu_x + u_{xxx} = 0)$ by the linear substitution of
variables:
\begin{equation}
\label{subs}
\begin{array}{l}
\delta = k /24, \, \tau = \delta \, t, \,
x = z - k \, t, \, u = a_2\,k\,\chi /6\,\delta\, a_1.
\end{array}
\end{equation}

Then soliton solution of (\ref{contin}) has a form
\begin{equation}
\label{sol}
\chi = \pm A\, sech^2 (B\,z \pm C\,t).
\end{equation}
Here $A = 12\,(k_2/k_3)\,B^2,\,\, C = B\,(k_1 + 4\,k_2\, B^2)$.
Solution of Eqn. \ref{sol} is a soliton with amplitude $A$ of either
compression $(A<0)$ or elongation $(A>0)$ propagating in the positive
or negative direction along the $z$--axis with velocity $v_{sol}=C/B$.
It describes a continuum set of one--parametric solitons, {\sl{i.e.}},
coefficients $A$ and $C$ can be expressed through the single
parameter, {\sl {e.g.}}, soliton half--width $w \approx 1.76/B$.

\begin{figure}
\centerline{\psfig{file=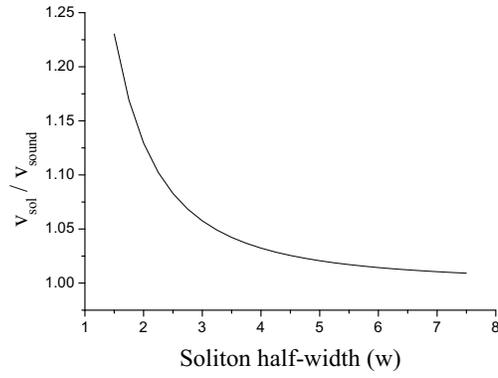,angle=0,width=8cm}}
\caption{The dependence of soliton velocity $v_{sol}$ on the soliton
half--width $w$. $v_{sound}$ is the longitudinal sound velocity.}
\label{soliton_fig2}
\end{figure}

Soliton velocity can be expressed as $v_{sol} = k_1 + 4\,k_2\,B^2 =
v_{sound}\,(1+0.52\,/w^2)$, where $v_{sound} = k_1 = \sqrt{a_1/M}$ --
is the longitudinal sound velocity ($ \approx 20$ km/sec). This value
is in good agreement with the value $20.35$ km/sec obtained in Ref.
\cite{sound}. The solution (Eqn. \ref{sol}) is, thus, the supersonic
soliton. For the solitons used extensively in our numerical
simulations (Sec. \ref{numerical}), the value $w=2$ was usually employed,
and $v_{sol} \approx 1.13\,v_{sound}$. Note, that the solution
(Eqn. \ref{sol}) has a form of solitary wave for {\it relative}
$z$--displacements. For an {\it absolute} $z$--coordinates it
transforms into a kink solution $\approx tanh(B\,z \pm C\,t).$ In the
limit $w \to \infty$ the soliton velocity tends to the sound velocity.
The dependence of $v_{sol}$ on $w$ is shown in Fig.
\ref{soliton_fig2}. Solitons with $w \le 1$ (less than the lattice
constant $\ell _0$) are highly unstable in discrete systems and,
therefore, this dependence is shown only for $w \ge 1.5$.

\begin{figure}
\centerline{\psfig{file=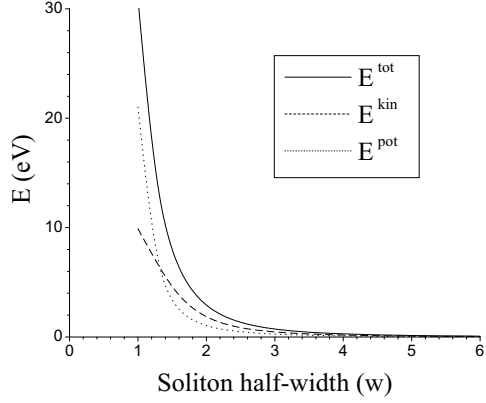,angle=0,width=8cm}}
\caption{The dependence of total soliton energy $(E^{tot})$ on soliton
half--width $w$ and individual contributions from kinetic $(E^{kin})$ and
potential $(E^{pot})$ energies.}
\label{soliton_fig3}
\end{figure}

The soliton in Eqn. \ref{sol} contains both kinetic and potential
energies and their contribution to the total energy of soliton
$(E^{tot})$ are roughly the same. The dependence of $(E^{tot})$ on $w$
as well as individual contributions from kinetic $(E^{kin})$ and
potential $(E^{pot})$ energies are shown in Fig. \ref{soliton_fig3}.

The solution (Eqn. \ref{sol}) was obtained using rather crude
approximations, neglecting higher order terms in the expansion
of Eqn. \ref{discrete} and radial displacements. In order to obtain a
more realistic soliton behavior, we next perform detailed numerical
simulations.

\section{Numerical simulation of solitons in CNT}
\label{numerical}

We present our results of the molecular dynamical simulation
using Brenner's many-body potential for the soliton evolution in the
($10,10$) CNT subject to different boundary conditions that include
free, rigid or capped ends. The findings for the $(6,6)$ CNT are
essentially the same.

Our CNT consisted of $1000 - 11000$ carbon atoms (number of layers
$N_{\ell } = 50 - 550$), with a nanotube length in the range $63 -
700$ \AA. The integration MD step was $0.35$ fs and Verlet--Beemann
algorithm was used to integrate the equations of motion. Initial
conditions were chosen in the form given by Eqn. \ref{sol} with
typical range of parameters values: $w=2 - 5, \, A=0.2 - 0.03$ \AA.

Solution (Eqn. \ref{sol}) describes the compression or elongation
soliton with variable amplitudes and velocities depending on the
half--width $w$, propagating on the CNT surface. This solitary
excitation is homogeneous with respect to the azimuthal angle $\Phi
$. As a first step in the numerical simulation we investigate the
soliton evolution in CNT with frozen radial displacements, {\sl{i.e.}}
by allowing only the $z$--displacements of the atoms.

\begin{figure}
\vspace*{-0.5cm}
\centerline{\psfig{file=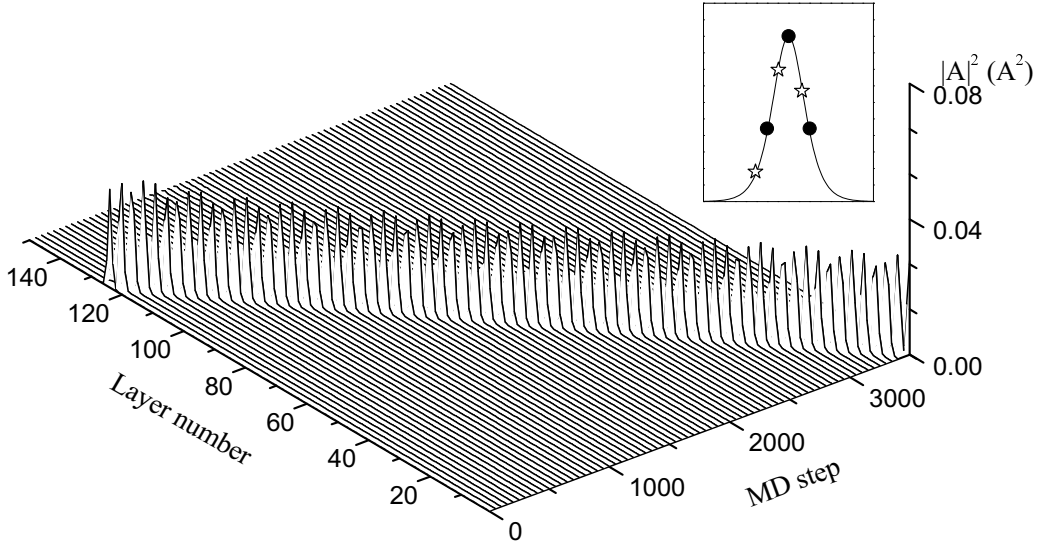,angle=0,width=15cm}}
\caption{Soliton evolution in the $(10,10)$ nanotube with 3\,000 atoms
(150 layers) with free ends. The soliton half--width is $w=2$ and $|A|^2 =
\chi ^2_n$ is the square of relative longitudinal displacements of the
$n$--th and $(n-1)$--th layers. The modulation of the soliton amplitude
in this figure is explained by the lattice discreteness and soliton
narrowness. Actual $\chi $ values are shown in insertion at two
different time instants (labeled by full circle and empty stars,
respectively). Only maximal values of $|A|^2$ reflect the soliton
stability.}
\label{soliton_fig4}
\end{figure}

The time evolution of a single soliton propagating along the $(10,10)$
nanotube is shown in Fig. \ref{soliton_fig4}. The soliton is rather
stable as it traverses the CNT length ($\approx 70$ nm) with only a
minimal decrease in amplitude. For CNT with cyclic boundary conditions
the soliton can travel much longer paths without suffering any
noticeable decrease in amplitude. The $|A|^2$ value, approximately
proportional to the elastic soliton energy in the harmonic
approximation, is plotted in Fig. \ref{soliton_fig4}. The wider is
the soliton, the greater its stability. In the limit of infinite CNT
diameter, Eqn. \ref{sol} describes a soliton propagating on the
graphite surface. Its behavior is likely to be similar to the soliton
in an isotropic hexagonal lattice with the Lennard--Jones inter-atomic
potential, propagating along the $[\overline{1}10]$
direction.\cite{Zolotaryuk}

While the soliton with half-width $w < 2$ interacts with open CNT ends
inelastically, a wider soliton reflects from the open ends without any
appreciable changes. Rigid CNT ends, on the other hand, are found to
preserve the soliton stability for any $w$.

\begin{figure}
\centerline{\psfig{file=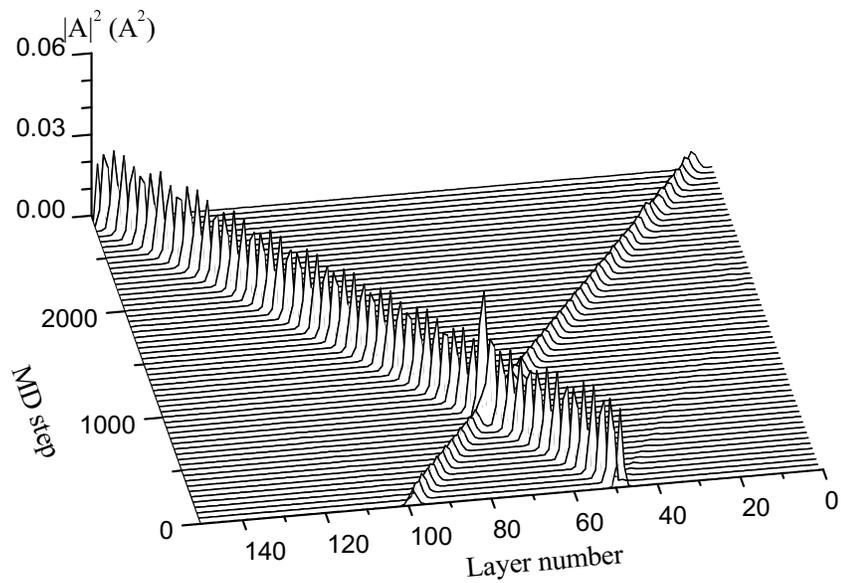,angle=0,width=15cm}}
\caption{Temporary evolution of two solitons with $w_1=2$ (right at
$t=0$) and $w_2=3$ (left at $t=0$) moving initially toward each
other. CNT and notations are the same as in Fig. \ref{soliton_fig4}.}
\label{soliton_fig5}
\end{figure}

The collision of two solitons with different parameters is illustrated
in Fig. \ref{soliton_fig5}. As seen in the figure, the solitons
collide and pass through each other without changing their profiles
and velocities, demonstrating their characteristic soliton feature.

Our numerical simulations performed at $T=300$~K showed no evidence of
the influence of thermal fluctuations on the soliton stability. Also,
there were no detectable soliton distortions in the presence of
defects (point mass defects, vibrationally excited bonds etc.) in the
CNT structure. Furthermore, no significant interaction with
low--intensity Raman--active tangential mode ($\sim 1585$
{cm}${}^{-1}$) was observed. Results of the MD simulations, therefore,
support the fact that the KdV equation is a good approximation for the
description of longitudinal nonlinear excitations -- solitons in CNTs
-- if radial displacements are frozen.

We have also performed MD simulations of soliton evolution
(propagation, reflection and collision between two solitons) in a 1D
lattice (a true 1D analogue of CNT) consisting of atoms interacting
through the potential described by Eqn. \ref{discrete}. Our results
are in full agreement with the findings for the CNTs, suggesting that
the neglect of higher order terms in the expansion of Brenner's
many-body potential (Eqn. \ref{Brenner}) is justified. The situation
changes, however, if the radial degrees of freedom are allowed. The
folding of the graphite sheet to form CNT results in the mixing of
longitudinal and radial modes. As a result, equations of motion for
these two modes now contain cross--terms, which can influence the
behavior of a longitudinal soliton.

The numerical simulation of the initial state evolution chosen in the
form of longitudinal soliton in CNT (Eqn. \ref{sol}) with {\it all}
allowed degrees of freedom shows the soliton to be less stable, with a
slow dissipation of the initial energy through the harmonic radial
oscillations. Results of analytical analysis and detailed numerical
investigations of the longitudinal--radial solitons will be presented
in a forthcoming paper.

\section{Interaction of solitary excitations with nanotube caps.}
\label{inter}

We next investigate the interaction of the solitons with nanotube
caps. Our results for the interaction of narrow ($w=2-4$) solitons
with non-local defects (free tube ends and caps) shows it to be
inelastic with the excitation energy being distributed approximately
evenly over all the bonds in this region. Interesting effects can be
expected in the case when the length scales of both the excitations
and defects are comparable. Towards this end, we perform a detailed
numerical investigation of interaction of excitations with the CNT
caps.

Solitons with $w=10$ carry too small an energy to induce any
noticeable effects in the cap. So we used solitary excitations with
other profiles. When a solitary excitation has a characteristic
length scale much greater than the lattice constant the nonlinear
effects can be ignored and can be treated within the harmonic
approximation. Furthermore, for long-wave excitations, the dispersion
effects in harmonic approximation are not significant. In this case,
an arbitrary excitation propagates along the harmonic chain with
speeds close to the sound velocity with only a minimal change in its
profile. We have examined the propagation of a longitudinal solitary
excitation using the parameters $\chi =
A\,sech^2[B\,(z-v_{sound}\,t)]$ with $w=10$ and $A=0.1$ \AA. This
excitation was found to be highly stable even when all degrees of
freedom were allowed and survived the reflection from free ends.

Same results were obtained for solitary waves with other analytical
profiles and similar widths and amplitudes. The other types of
excitations studied included an excitation of the Gaussian profile
$\chi = \displaystyle\frac{A}{\sqrt{2\pi w}}\,
exp\left[-\displaystyle
\frac{\left(z-v_{sound}\,t\right)^2}{2\,w^2}\right]$, also with $w=10$
and $A=0.1$ \AA.

Although the soliton propagation is accompanied by a highly coherent
and predominantly longitudinal displacements of the atoms on the
cylindrical surface of the CNT, the atoms on the cap behave quite
differently after the solitary wave -- cap collision.
When the solitary excitation reaches the cap, there is an
accumulation of energy at the cap. Part of the total energy of the
solitary wave is likely to be dissipated and/or accumulated on
structural defects. We consider an example in which solitary wave
interacts with non symmetrical caps.

Our simulation of soliton interaction with the nanotube cap consisted
of the following sequences: (A) The incoming excitation causes the cap
to ``inflate'' with the cap attaining nearly a spherical shape. At
this instant the potential energy of the cap attains a maximum value,
although the energy distribution in the bonds are unequal due to the
asymmetry in the cap. Atoms in the tip of the cap are displaced by
$\approx 0.3 $ {\AA} in the positive $z$--direction; (B) The recoil of
the elastic energy of the cap causes atoms to move in the opposite
direction and at some instant the kinetic energy attains a maximum
value. The velocities of atoms are all predominantly directed along
the $z$--axis although they differ significantly from each other in
their magnitudes; (C) Atoms at the cap continue to be propelled by
inertia and the elastic potential energy attains a maximum again. At
this instant, the cap acquires regions with negative curvature, and
the displacement of the tip of the cap from its equilibrium position
is $\approx -0.8$ \AA; (D) After this the atoms in the cap start to
move in the opposite direction toward the equilibrium state and
second, but less intense, maximum in the kinetic energy is observed.

The nanotube cap continues to oscillate with rapidly decreasing
amplitude, the energy being dissipated in heating the cap and the CNT
``body''. Note the large total amplitude of the tip displacements
($>1.0$ \AA). A striking feature of this preliminary results observed
in MD simulations is the inhomogeneity of the energy distribution in
the cap after the solitary wave -- cap collision. We next investigate
the cap dynamics and energy distribution over atoms in more details.

\begin{figure}
\centerline{\psfig{file=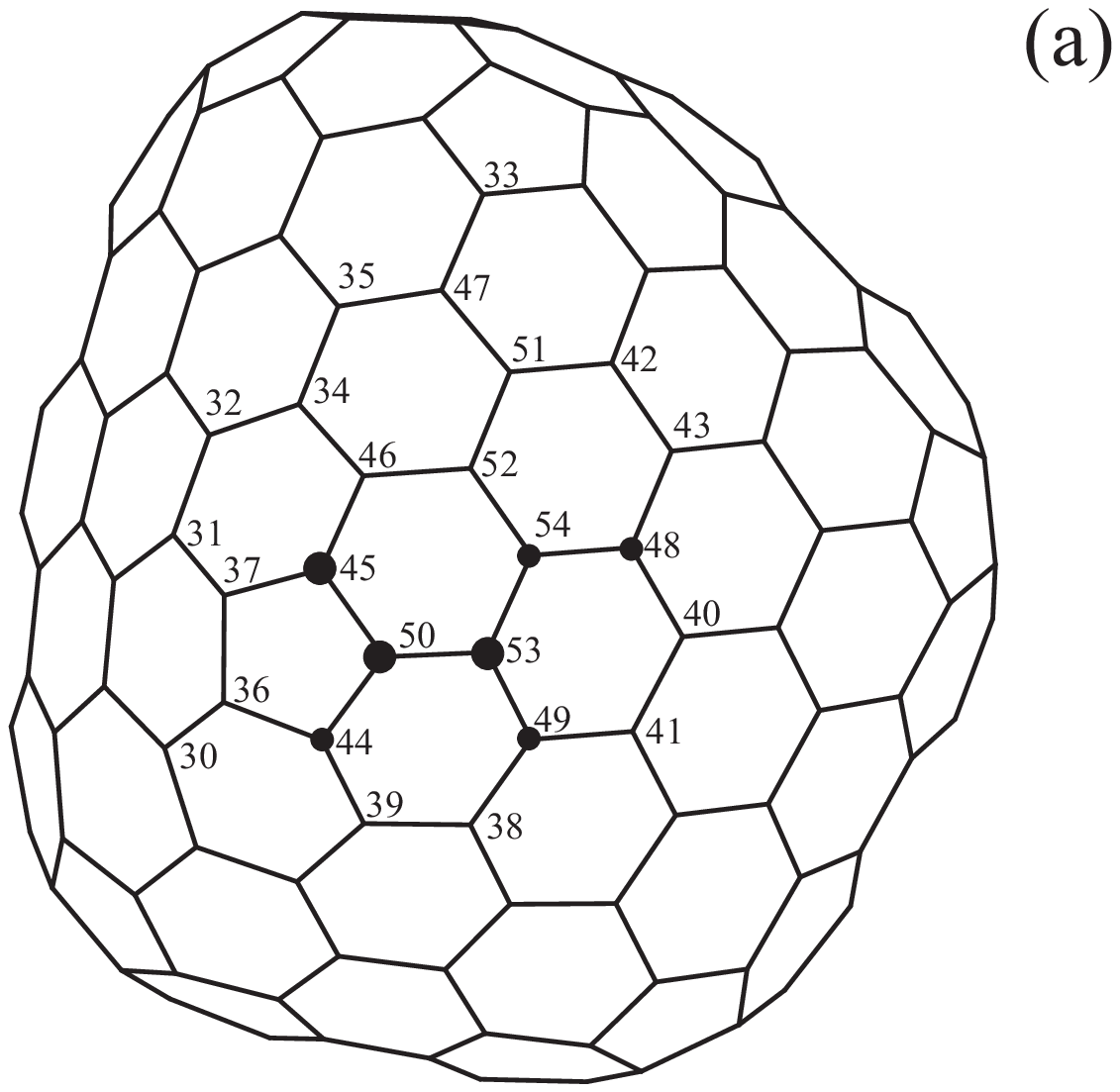,angle=0,width=8cm}}
\centerline{\psfig{file=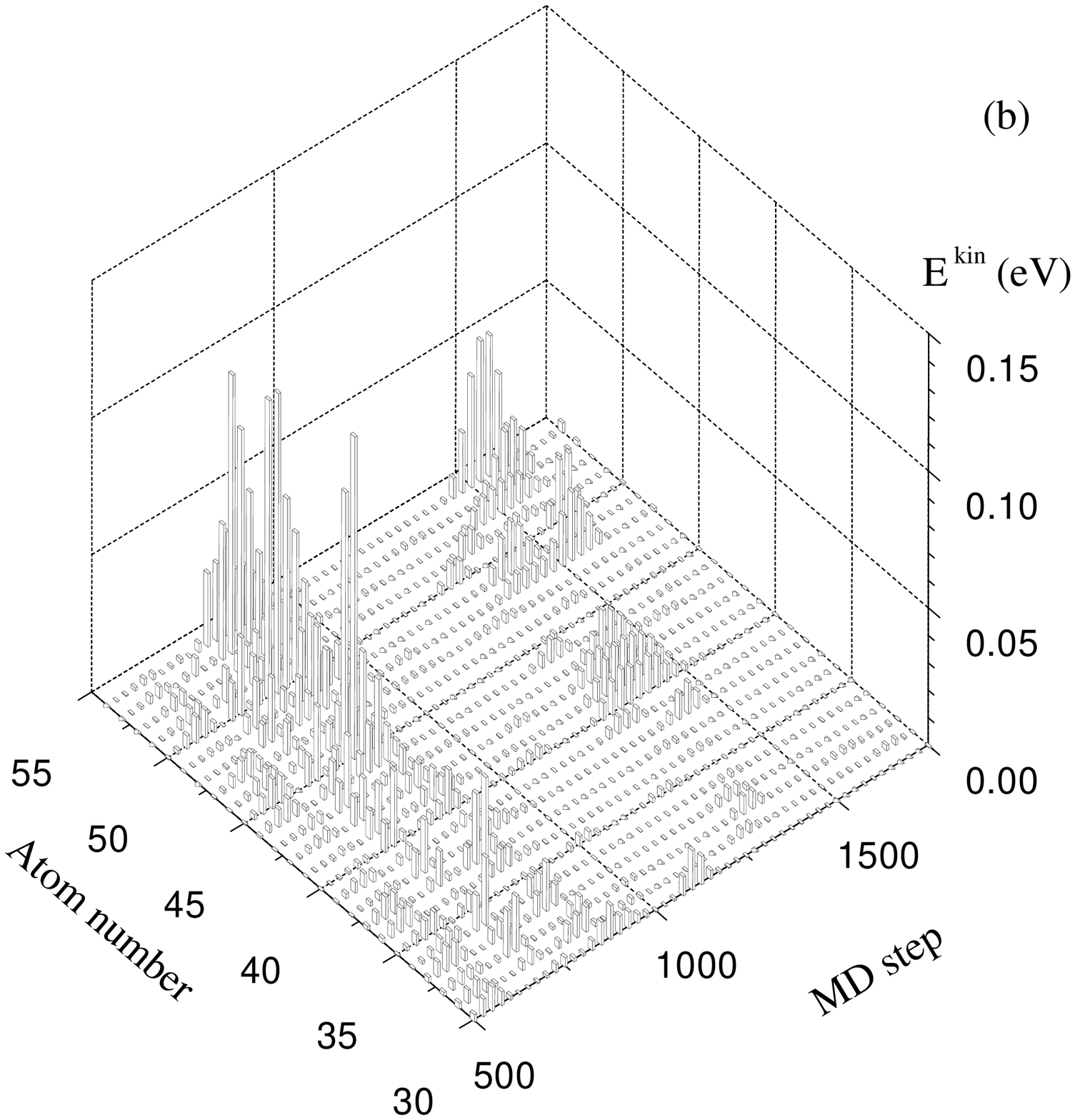,angle=0,width=15cm}}
\caption{(a) Schlegel diagram and the atomic numbering of
non--symmetrical (10,10) CNT cap; (b) temporary evolution of the
kinetic energy distribution over atoms in the cap. Initial condition
-- solitary excitation with $w=10$ moves from the tube to the cap and
at $t \approx 600$ MD step reaches the cap. The maximum energy
concentration is attained at $t \approx 800$ MD steps.}
\label{soliton_fig6}
\end{figure}

A typical time dependence of the kinetic energy distribution over
atoms in the $(10,10)$ cap is shown in Fig. \ref{soliton_fig6}b as a
histogram. Schlegel diagram of this cap is shown in Fig.
\ref{soliton_fig6}a. This cap was chosen from a full list of $9\,342$
topologically different caps with isolated pentagons generated
previously.\cite{our} Atom numbering is common to both Figs.
\ref{soliton_fig6}a and \ref{soliton_fig6}b, and atoms with numbers
less than 30 (in the cap base) are not shown as they have lower
($<0.02$ eV) excitation levels. Two ``waves'' of excitations are
visible in this Figure: one at $t=750 - 800$ MD steps (stage (B)) and
the other with lower excitation level -- at $1300 - 1600$ MD steps
(stage (D)). The atoms with energy $\geq 0.1$ eV are drawn in large
circles in Fig. \ref{soliton_fig6}a; atoms with energy $\geq 0.05$ eV
-- in medium circles. An excitation area is localized and bears a
larger portion of total cap energy. One can see that the kinetic
energy is concentrated mainly at the tip of the cap and irrespective
of the location of pentagons.

\begin{figure}
\centerline{\psfig{file=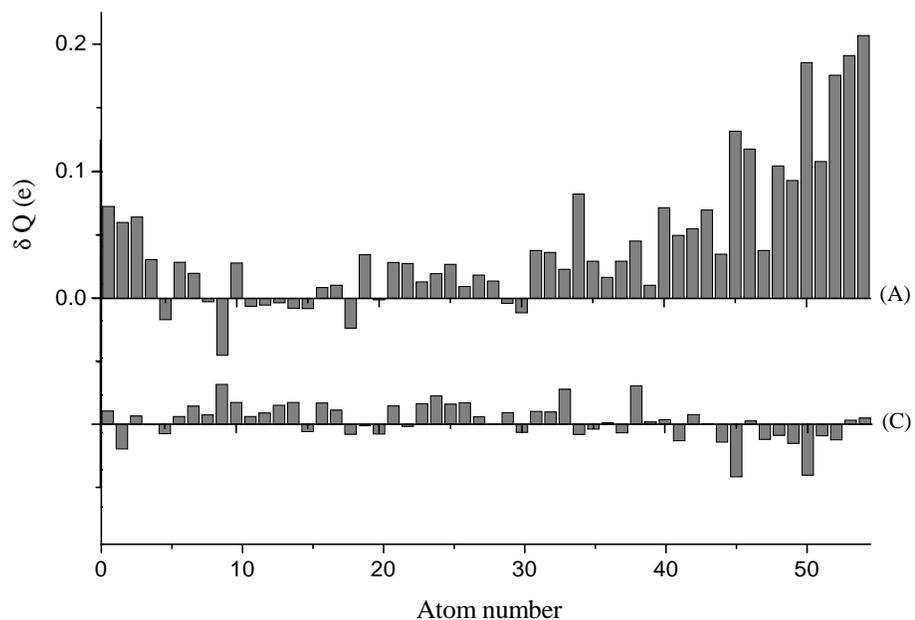,angle=0,width=15cm}}
\caption{Variation of the partial Mulliken charges at the cap
at different time instants {\sl vs.} charges of the relaxed cap.
Atomic numbering is the same as in Fig.\ref{soliton_fig6}.
(A) -- largest ``positive'' expansion of the cap
(stage (A)); (C) -- ``negative'' shrinkage (stage (C)).}
\label{soliton_fig7}
\end{figure}

We denote this effect of energy concentration as ``Tsunami effect''
(named for the effect explaining ocean waves coming to a beach)
because initially long--wave and low--amplitude excitation is
concentrated into a sharp impulse of energy if the conditions for the
solitary wave propagation are changed in a special manner. It was
found that this effect depends on the cap structure and is more
pronounced in less symmetrical caps. The effect is observed if the
cap radius is comparable with solitary wave half--width $w$ (in the
case shown in Fig. \ref{soliton_fig7} for $w=10$). Otherwise, the
Tsunami effect is less prominent. Analogously similar phenomena with
varying degrees of perfections were observed for other defects (kinks
and bends) in CNT.

We have also calculated electronic properties of the capped CNTs
during the solitary wave -- cap collision using a generalized tight
binding scheme of Menon and Subbaswamy.\cite{Menon} The variation of
the partial Mulliken charge distribution obtained using this scheme is
shown in Fig. \ref{soliton_fig7}. These calculations were performed
for time instants corresponding to the maximal cap distortions (stages
(A) and (C)). The most significant change, as expected, is observed
for atoms at the tip of the cap (indicated by large number lables in
Fig. \ref{soliton_fig7}) where perturbation in the structure is a
maximum.

\begin{figure}
\centerline{\psfig{file=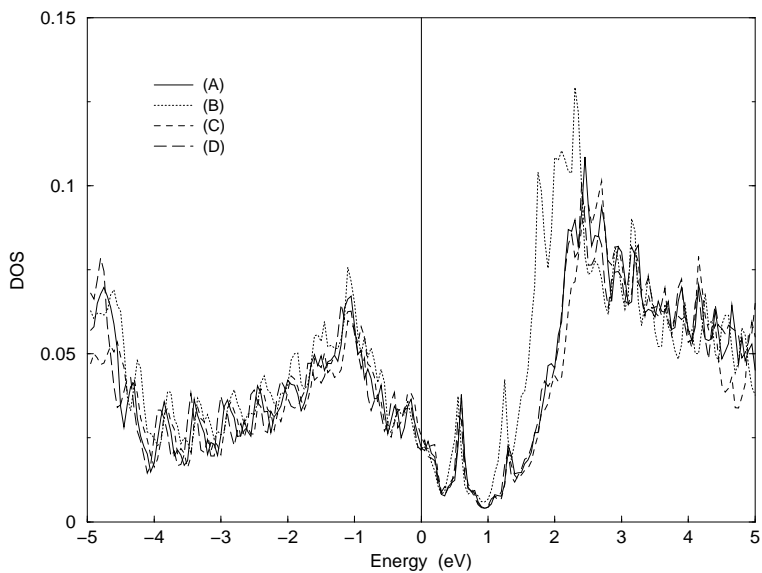,angle=-90,width=10cm}}
\caption{Variation of the densities of states averaged over the cap
atoms at different instances.}
\label{soliton_fig8}
\end{figure}

The local electron density of states (DOS) averaged over atoms at the
cap at stages (A) through (C) of solitary wave -- cap collision are
shown in Fig. \ref{soliton_fig8}. The DOS is obtained from the
Green's function G(E)=[ES-H+i$\delta]^{-1}$, where H is the
generalized tight binding Hamiltonian, S the overlap\cite{Menon}, and
$\delta=0.05$ eV. Note the rather large variations in the DOS.

\section{Conclusion}
\label{concl}

We have, thus, demonstrated analytically and using numerical
simulations that the KdV solitons are reasonably good approximations
for the description of nonlinear excitations in CNTs. The soliton
stability increases with the CNT diameter and, in the limit $R \to
\infty $, the solution (Eqn. \ref{sol}) describes the nonlinear
excitations in graphite. Our numerical simulations of the evolution
of the initial longitudinal soliton confirms its high stability.

We have also performed numerical simulations of the interaction of
solitary excitations (different from solitons) with defects in CNTs,
especially with caps (the Tsunami effect). Our simulations show many
new fascinating features in the dynamics of CNTs. The collision is
found to be partially inelastic, and the cap excitation is highly
inhomogeneous. Interestingly, at some instances the energy in a few
bonds considerably exceeds the averaged energy of solitary
excitation. This process can provide extra dynamical contribution to
the recently discovered phenomenon of the enhanced reactivity of
defect sites in CNTs termed ``kinky
chemistry''.\cite{Srivastava,Ausman}

The issues concerning ways of solitons excitation in real CNTs and
their possible contribution to various chemical and physical phenomena
are not yet clear. In reality, the generation of the highly coherent
ultra short longitudinal displacements of atoms in the CNT seems
hardly probable. This, however, is not the case for the long wave
solitary excitations. In this connection we would like to point out
recent results on the opto--mechanical effect in
CNT.\cite{Zhang,Poncharal} Conceivably a short flash of light could
trigger the CNT excitation in such a manner that the solitary
excitations will be generated. They can also be generated due to
external factors such as electron or ion impacts, stress release and
other mechanisms. If solitary excitations do exist in CNTs in
reasonable ``concentration'', then they can be identified through
their contribution to some detectable CNT properties. A possible
example is the Tsunami effect when the energy, smoothly distributed in
the large scale excitation of an arbitrary profile, can be
concentrated on a few bonds of a defect with considerable energy
excess. The greater the distortions in CNTs, the greater is this
influence. It can also be a way for a ``self--healing'' of CNTs when a
rearrangement of structural imperfection is activated by solitary
excitations. This effect also can promote chemical reactions such as
$C_2$ incorporation into nanotube caps.

The high specific heat of a rope of SWNT observed\cite{Mizel} can be
partially explained by solitary excitations generated in parallel with
phonons. Heat transfer in CNTs is another
example.\cite{Yi,Hone,Berber} Emission of short electric field pulses,
when solitary wave inelastically interacts with the non--symmetrical
caps or other defect sites, can be yet another example. Our results of
the calculation of variation in charge distribution and DOS confirm
large deviation of the electronic properties from their equilibrium
values at the solitary excitation -- cap collision. The process of
charge distribution variation under dynamical excitation is probably
inverse to the opto--mechanical phenomenon in CNTs.\cite{Zhang},
\cite{Poncharal} Nanotechnology is one area where the
mechano--electric property of CNT can be used.

We had initially planned to title our paper ``Solitons in carbon
nanotubes'', but after the preparation of this manuscript we became
aware of a very recent work on soliton by Chamon\cite{Chamon} with the
same title. In this paper the spontaneous lattice distortions are
investigated similar to the case of polyacetylene. The solitons are
topological domain walls separating different symmetry-broken vacua
with different Kekule bond-alternation structures. Note that the
solitons discussed by Chamon and those found in the present work have
different nature: we have considered solitons formed due to the
elastic and non-linear properties of CNT, while Chamon solution is a
topological soliton.

The authors (T.A., O.G., and G.V.) thank the Russian Foundation for
the Basic Researches (Project 98--03--32218a ) for the partial
financial support. T.A. and O.G. are also indebted to RFBR Project
00--15--97334. Stimulating discussions with A.A.Ovchinnikov and
A.V.Zabrodin are gratefully acknowledged. M.M. acknowledges support
through grants by the NSF (No. 99-07463, MRSEC Program under Grant No.
DMR-9809686), DEPSCoR (No. 99-63231 and No. 99-63232), DOE Grant (No.
00-63857), and the University of Kentucky Center for Computational
Sciences.

\newpage

\newpage

\newpage

\newpage

\newpage

\newpage

\begin{table}

\caption{ Coefficients (in $eV$) in the expansion of
Brenner's potential (Eqn. \ref{Bren_tay3}) and the
equilibrium distance between layers $\ell _0$ (in \AA) for CNTs of
different diameters.}

\begin{tabular}{|c|c|c|c|c|}
& $(5,5)$ & $(10,10)$& $(15,15)$& $(20,20)$\\
\hline
$a_1$ & 102.80 & 102.76 & 102.70 & 102.66 \\
$a_2$ & 296.46 & 296.22 & 295.86 & 294.39 \\
\hline
$b_1$ & 11.80 & 2.98 & 1.32 & 0.76 \\
$b_2$ & 0.34 & 0.10 & 0.04 & 0.00 \\
$b_3$ & 9.06 & 4.59 & 3.09 & 2.31 \\
$b_4$ & 18.42 & 2.40 & 0.72 & 0.30 \\
\hline
$c_1$ & 12.24 & 6.16 & 4.14 & 3.10 \\
$c_2$ & 71.34 & 35.82 & 24.06 & 17.94 \\
$c_3$ & 152.40 & 154.56 & 154.20 & 153.66 \\
\hline
$\ell _0$ & 1.2601 & 1.2573 & 1.2568 & 1.2565
\end{tabular}
\end{table}

\end{document}